\documentclass{svproc}
\usepackage{url}

\usepackage{graphicx,color}
\usepackage{hyperref}

\begin{document}
\mainmatter              
\title{Light Nuclei from Lattice QCD:
\\
Spectrum, Structure and Reactions}
\titlerunning{Light Nuclei from Lattice QCD}  
%
\author{Zohreh Davoudi~\inst{1},\inst{2}
\\
\emph{For the NPLQCD collaboration}}
\authorrunning{Zohreh Davoudi} 
%
%
\institute{Maryland Center for Fundamental Physics and Department of Physics, 
University of Maryland, College Park, MD 20742, USA\\
\and
RIKEN Center for Accelerator-based Sciences,
Wako 351-0198, Japan
\email{davoudi@umd.edu}
}

\maketitle              

\begin{abstract}
Lattice Quantum Chromodynamics (LQCD) studies of light nuclei have entered an era when first results on structure and reaction properties of light nuclei have emerged in recent years, complementing existing results on their lowest-lying spectra. Although in these preliminary studies the quark masses are still set to larger than the physical values, a few results at the physical point can still be deduced from simple extrapolations in the quark masses. The progress paves the road towards obtaining several important quantities in nuclear physics, such as nuclear forces and nuclear matrix elements relevant for $pp$ fusion, single and double-$\beta$ decay processes, neutrino-nucleus scattering, searches for CP violation, nuclear response in direct dark-matter detection experiments, as well as gluonic structure of nuclei for an Electron-Ion Collider (EIC) program. Some of the recent developments, the results obtained, and the outlook of the field will be briefly reviewed in this talk, with a focus on results obtained by the Nuclear Physics From LQCD (NPLQCD) collaboration.

\end{abstract}
\section{One the Goals and Impact of a LQCD Program for Nuclear Physics
\label{sec:goals}}
\begin{figure}[t!]
\includegraphics[scale=0.265]{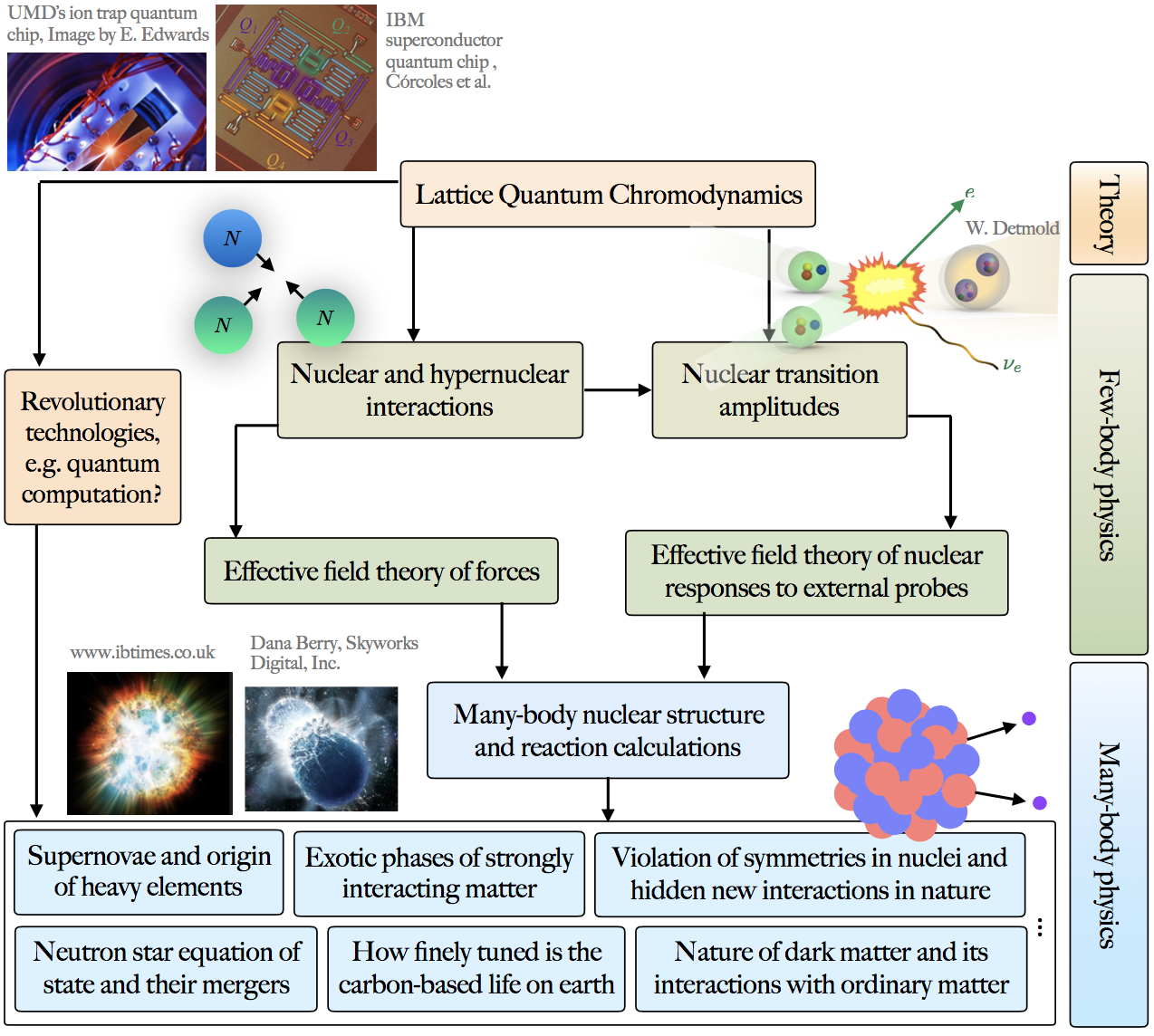}
\caption[.]{A roadmap illustrating a systematic path from QCD to addressing grand-challenge problems in nuclear physics.}
\label{fig:roadmap}
\end{figure}

\noindent
The standard approach in nuclear structure and reaction theory has shifted from relying on phenomenological nuclear potentials to studies based on nuclear effective field theories (EFTs), hence providing a systematic way to assess uncertainties of a calculation. In order for this program to succeed, not only the nuclear EFTs must offer a valid power counting with convergent and renormalization-scale independent results, but also their multitude of low-energy coefficients (LECs) must be fit to experiment, so the EFTs can acquire a predictive power. In situations where experimental data are scarce or nonexistent, such as multi-neutron and hyperon-nucleon interactions, nuclear effects in the response of a nucleus to external probes, or nuclear matrix elements for the neutrinoless double-$\beta$ decay of a nucleus, studies based on the underlying theory of quantum chromodynamics (QCD) are essential. Reliable predictions for a number of grand-challenge problems in nuclear physics  (a few examples of which are enumerated in the chart in Fig.~\ref{fig:roadmap}) will benefit from a coherent program in nuclear theory, in which the input from the underlying theory of QCD in the few-body sector provides the stepping stone for a nuclear many-body study based upon the constrained EFTs. A roadmap of this program is depicted in Fig.~\ref{fig:roadmap}.

The only reliable method that enables QCD determination of observables in nuclear physics is LQCD, a method that relies on Monte Carlo sampling of the quantum fields in QCD, and provides $n$-point correlation functions obtained in a finite discretized Euclidean spacetime. Physical observables can be obtained in a systematic way using various extrapolations, or in the case of scattering amplitudes and transition rates, through mappings between the finite and infinite-volume theory. Heroic effort has been devoted in recent years to studies of multi-nucleon systems,  considering the great computational complexity of these studies, and impressive progress has been made. While this short review can not do justice to the wealth of the results obtained in this area, I will focus on selected results by the NPLQCD collaboration on hadronic interactions, nuclear structure and nuclear reactions from LQCD. For a recent review of multi-nucleon physics from LQCD, see Refs.~\cite{Davoudi:2017ddj}.
\section{Hadronic Interactions
\label{sec:forces}}
\begin{figure}[t!]
\includegraphics[scale=0.525]{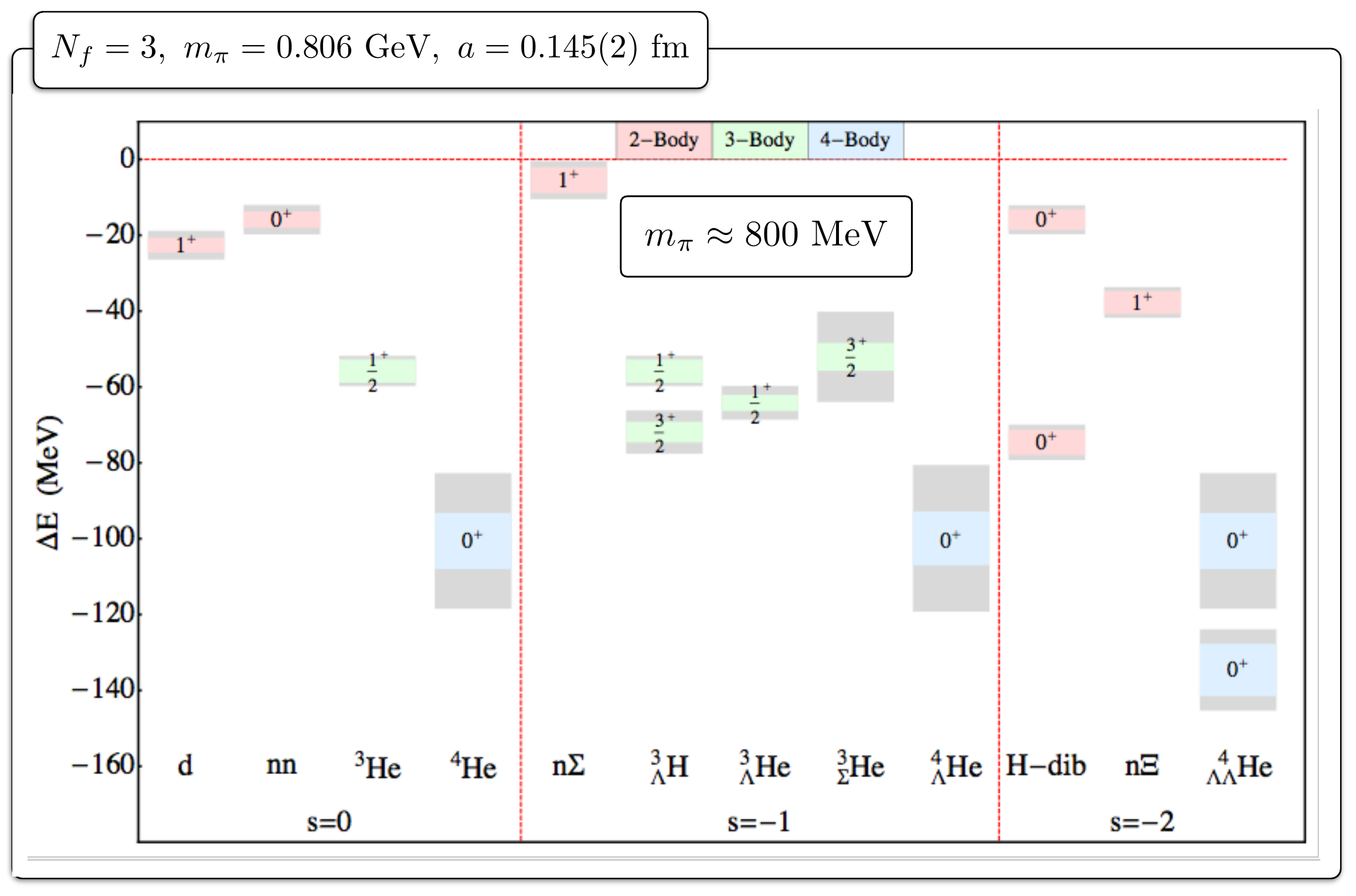}
\caption[.]{The lowest-lying spectra of light nuclei and hypernuclei from LQCD obtained at larger-than-physical values of the quark masses~\cite{Beane:2012vq}. The figure and all subsequent figures courtesy of the NPLQCD collaboration.}
\label{fig:nuclei}
\end{figure}
\noindent
Constraining nuclear and hypernuclear forces remains a central component of research in nuclear physics. This effort complements experiments on neutron-rich isotopes, and provides the input to research on the nature of dense matter in astrophysical environments. A milestone for nuclear physics from LQCD was reached in 2012 when the emergence of light nuclei and hypernuclei from QCD was demonstrated in Ref.~\cite{Beane:2012vq}, albeit at larger-than-physical values of the quark masses, Fig.~\ref{fig:nuclei}. This was enabled by algorithmic advances in forming nuclear correlation functions based on Refs.~\cite{Detmold:2010au,Detmold:2012eu} and the availability of computational resources. Such spectral studies at closer-to-physical values of the quark masses have since been conducted and appear promising. Further, a leap in the application of LQCD to nuclear physics was the realization~\cite{Beane:2003da} that large unnatural scattering lengths in two-nucleon systems is not an impediment in applying the powerful Lu\"scher's method~\cite{Luscher:1986pf} -- a method that turns the finite-volume spectra obtained from LQCD to scattering amplitudes in the two- (and in recent extensions of the method to three-) body scattering amplitudes.

\begin{figure}[t!]
\includegraphics[scale=0.571]{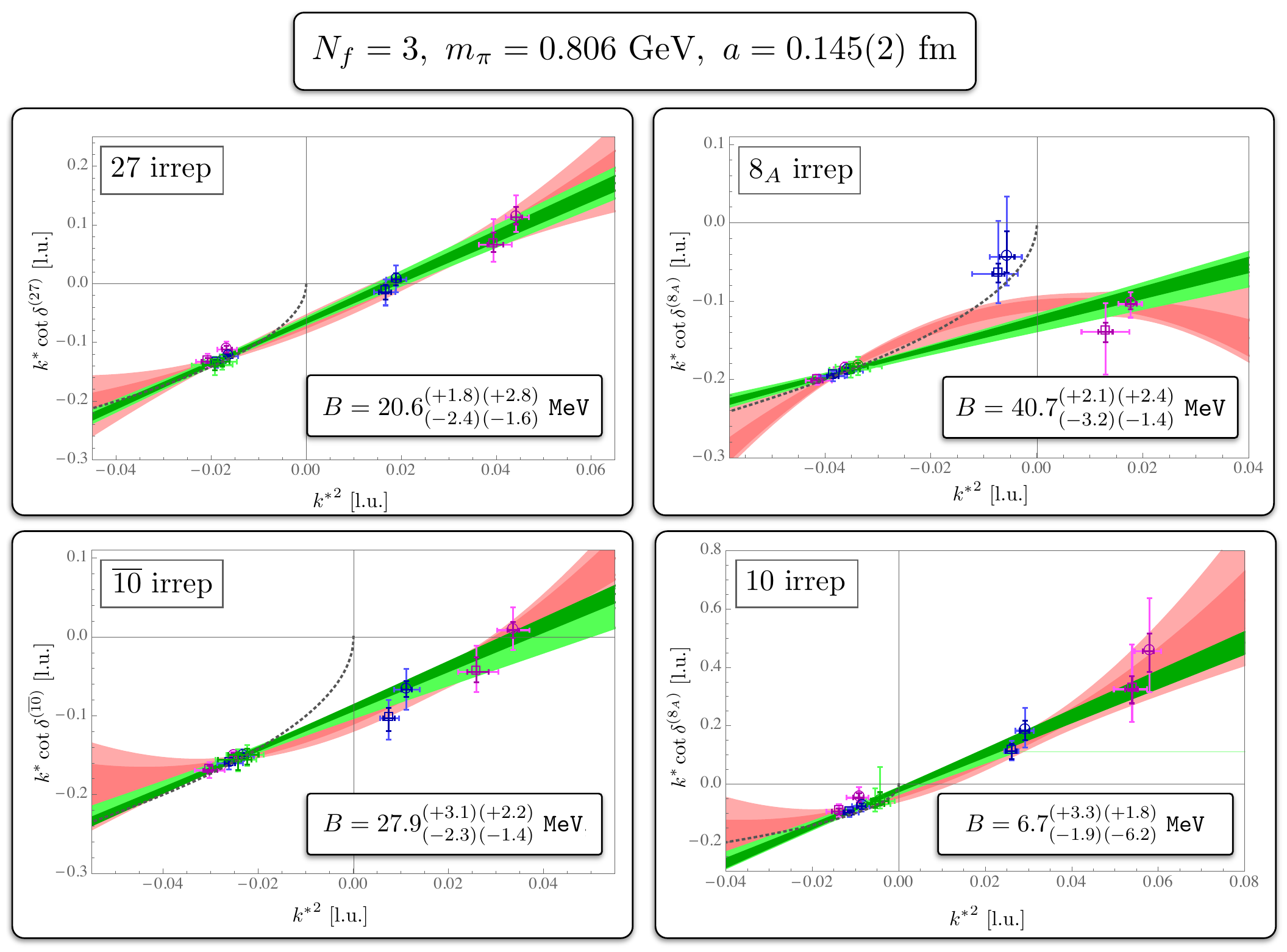}
\caption[.]{Low-energy scattering phase shifts of various two octet-baryon channels along with the binding energy of the lowest-lying states, obtained with LQCD at larger-than-physical values of the quark masses~\cite{Wagman:2017tmp}. }
\label{fig:bb}
\end{figure}

A recent example of such application is shown in Fig.~\ref{fig:bb}, in which the SU(3) flavor-symmetric s-wave scattering phase shifts and the ground-state binding energies in four different scattering channels, corresponding to representative flavor channels $NN({^1}S_0)$, $NN({^3}S_1)$, $N\Sigma({^3}S_1)$ and $N\Xi({^3}S_1)$, were constrained, albeit at larger-than-physical values of the quark masses~\cite{Wagman:2017tmp}. An interesting finding, arrived at by the observation of nearly identical scattering lengths and effective ranges in the four SU(3) flavor-symmetric channels, is that the low-energy interactions among two octet baryons exhibit not only a SU(6) spin-flavor symmetry that is argued to exist in QCD in the limit of a large number of colors~\cite{Kaplan:1995yg}, but also an extended SU(16) symmetry, which is now conjectured to be in place to minimize the entangling power of the S-matrix at low energies~\cite{Beane:2018oxh}. Further, this study demonstrates the matching between LQCD output and the EFT LECs, a program that can enable studies of larger systems of nucleons currently not accessible directly with LQCD.
%

\section{Nuclear Structure
\label{sec:structure}}
\begin{figure}[t!]
\includegraphics[scale=0.475]{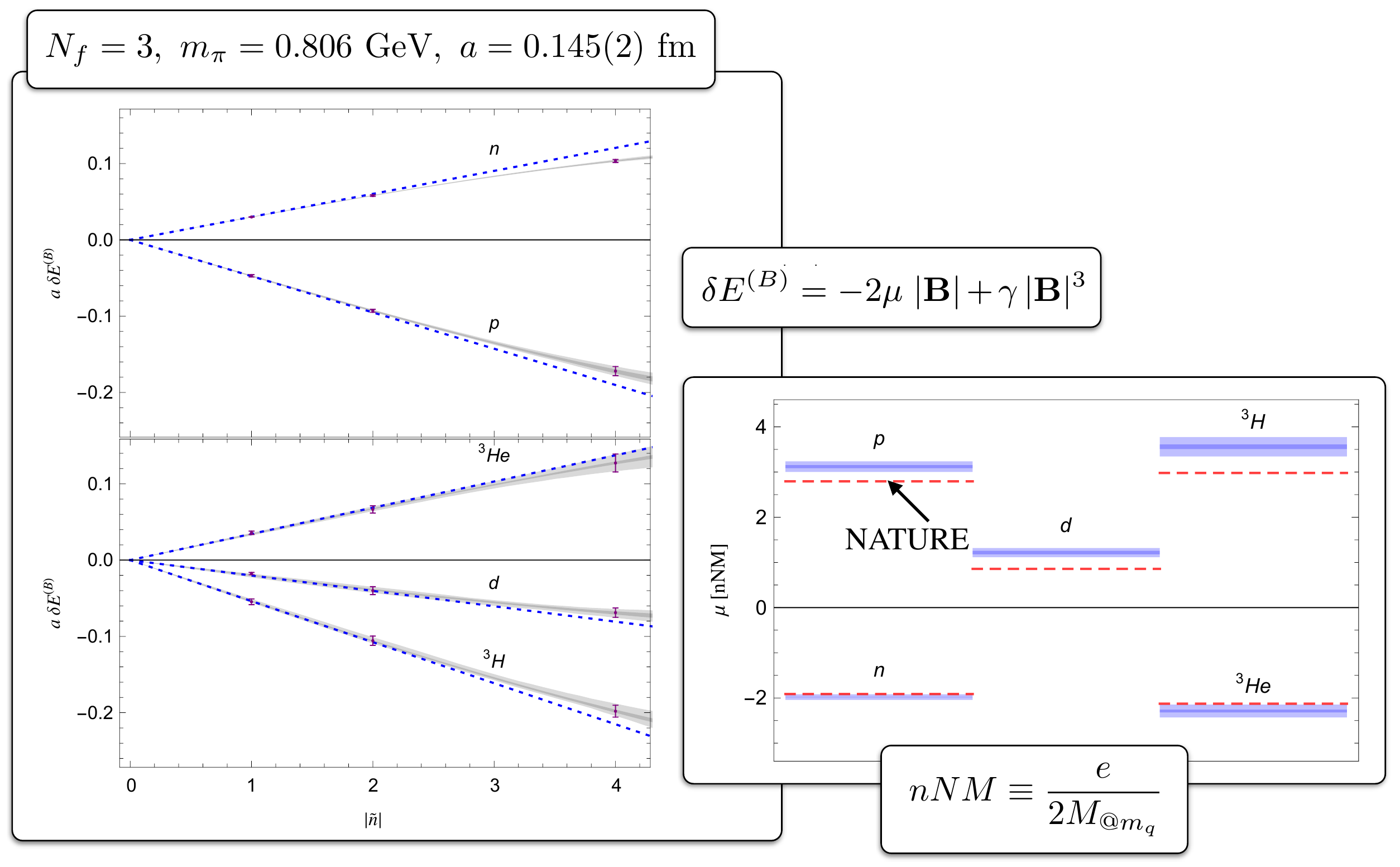}
\caption[.]{The shift in the energy of select light nuclei, $\delta E^{(B)}$, in a background magnetic field $\mathbf{B}$ (left) and the extracted magnetic moments, $\mu$, from LQCD at larger-than-physical values of the quark masses (right)~\cite{Beane:2014ora}.}
\label{fig:magmoment}
\end{figure}

\noindent
Investigations into the structure of hadrons and nuclei aim to provide further insight into the nature of strong dynamics. They are further essential in interpreting the outcome of high-energy collider experiments by providing a more accurate picture of the internal structure of the colliding protons or heavy ions. Like in experiment, certain electromagnetic (EM) properties of hadrons, such as magnetic moment, electric and magnetic polarizabilities and charge radii, can be deduced in a LQCD calculation from the response of the hadron to external EM fields. Such studies have been extended to light nuclei in recent years. As is shown in Fig.~\ref{fig:magmoment}, the shift in the lowest finite-volume energy of proton, neutron, deuteron, $^3$He and $^3$H in an external magnetic field are used to deduce their magnetic moment, albeit at a larger-than-physical value of the quark masses~\cite{Beane:2014ora}. When expressed in units of natural nuclear Magneton defined with the mass of the nucleon/nuclei at the corresponding value of the quark masses, they are surprisingly close to their values in nature, suggesting that much of the quark-mass dependence of the magnetic moment is captured by the quark-mass dependence of the mass. Additionally, it is observed that as in nature, nuclei at such large values of the quark masses still appear to behave as a collection of the nucleons, i.e., they can be described by a shell-model picture.
\begin{figure}[t!]
\includegraphics[scale=0.495]{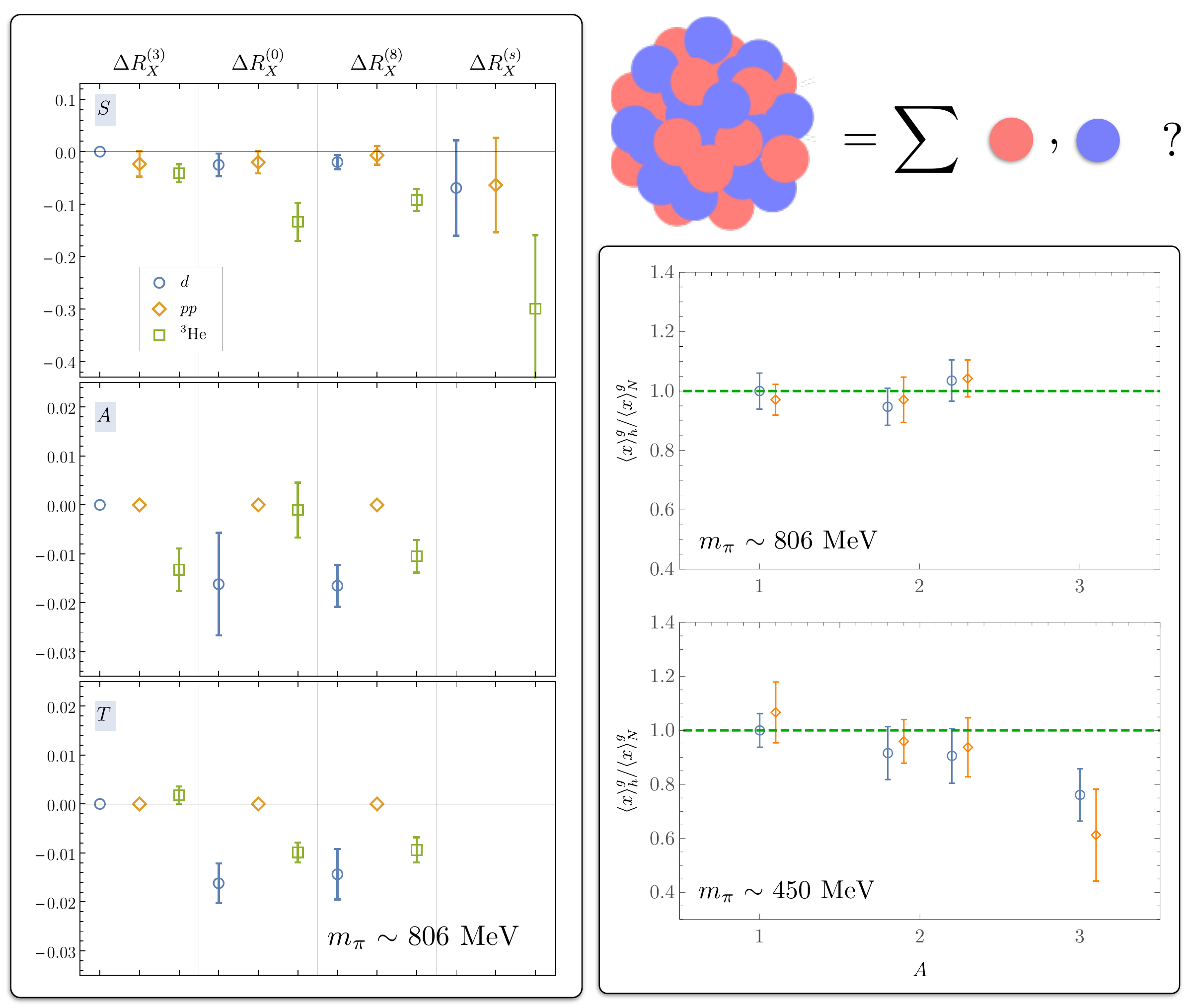}
\caption[.]{Depicted in left is the deviation of the matrix element $\langle h | \bar{q}\Gamma q | h \rangle$ from that in a noninteracting model of nucleons. $\bar{q}\Gamma q$ denotes scalar, axial and tensor quark bilinear currents and $h=\{d,pp,{^3}\rm{He}\}$. Depicted in right is the ratio of the gluon momentum fraction in select light nuclei to that of the single nucleon, obtained from LQCD  at larger-than-physical values of the quark masses~\cite{Chang:2017eiq}. In the right panel, blue and orange colors correspond to two different sink operators in forming the correlation functions. For a detailed description of quantities, see Refs.~\cite{Chang:2017eiq,Winter:2017bfs}. 
}
\label{fig:structure}
\end{figure}

A further motivation for a nuclear structure program from LQCD is in supporting experiments in Fundamental Symmetries and Searches for New Physics. For example, LQCD studies of the matrix elements of scalar, axial and tensor currents in light nuclei can determine how significant nuclear effects (those arising from the fact that a nucleus is more than a collection of nearly noninteracting nucleons) are for current and future searches for CP violation in nuclei, in the single and double-$\beta$ decay of a large nucleus and in the direct searches for dark matter candidates using heavy isotope as targets. This is enabled through matching the LQCD results in the few-body sector to the corresponding EFT description of these processes, a process that can constrain unknown two and multi-nucleon short-distance effective couplings of the EFT, see Fig.~\ref{fig:roadmap}. A first LQCD study of scalar, axial and tensor quark-bilinear currents in light nuclei was performed in Ref.~\cite{Chang:2017eiq} at a large value of the quark masses (see the left panel of Fig.~\ref{fig:structure}), and found nonnegligible nuclear effects in the scalar matrix element. If this conclusion persists at the physical values of the quark masses, significant nuclear effects may need to be accounted for in obtaining the cross section of nuclear targets with dark-matter candidates in scalar portals.

Finally, LQCD enables gluonic probes of nuclear structure, promising a growing program that can provide the theoretical support for an EIC in upcoming years. Investigations into the role of the gluons in the structure of single hadrons have reached significant milestones in recent years, while in the case of the nuclei, this effort has just been started. The first LQCD determination of select gluonic observables in light nuclei was reported in Ref.~\cite{Winter:2017bfs}, albeit at two unphysically large values of the quark masses, in search for a gluonic analog of the EMC effect, see the right panel of Fig.~\ref{fig:structure}. These studies demand significant computational resources but are planned to be improved, in scope and precision, in upcoming years. 
\section{Nuclear Reactions
\label{sec:reactions}}
\begin{figure}[t!]
\includegraphics[scale=0.45]{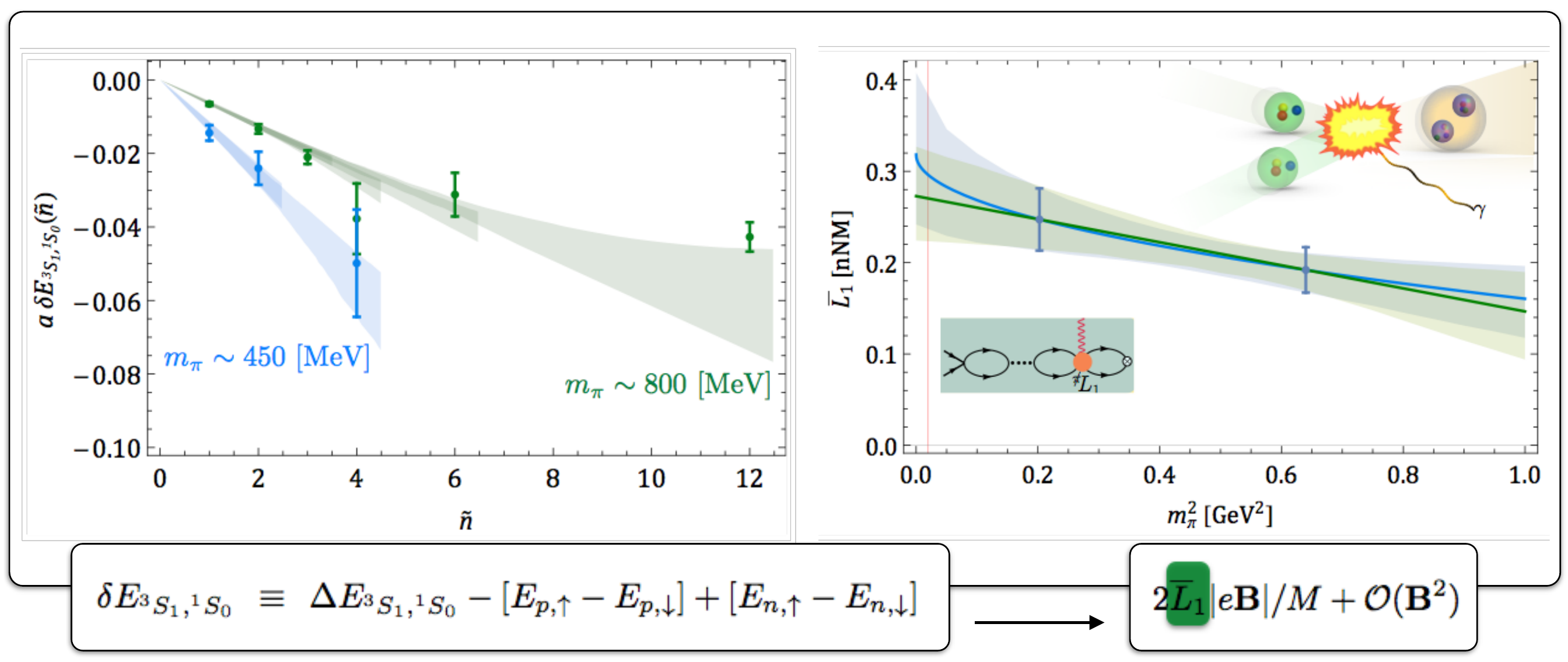}
\caption[.]{The background magnetic field breaks the (near) degeneracy of two-nucleon systems in a LQCD calculation (left). This gives access to the nuclear matrix element for the M1 magnetic transition in the radiative capture process $np \to d\gamma$, and constrains the two-nucleon short-distance coupling of the pionless EFT, $L_1$, (right)~\cite{Beane:2015yha}. }
\label{fig:npdgamma}
\end{figure}
\begin{figure}[t!]
\includegraphics[scale=0.45]{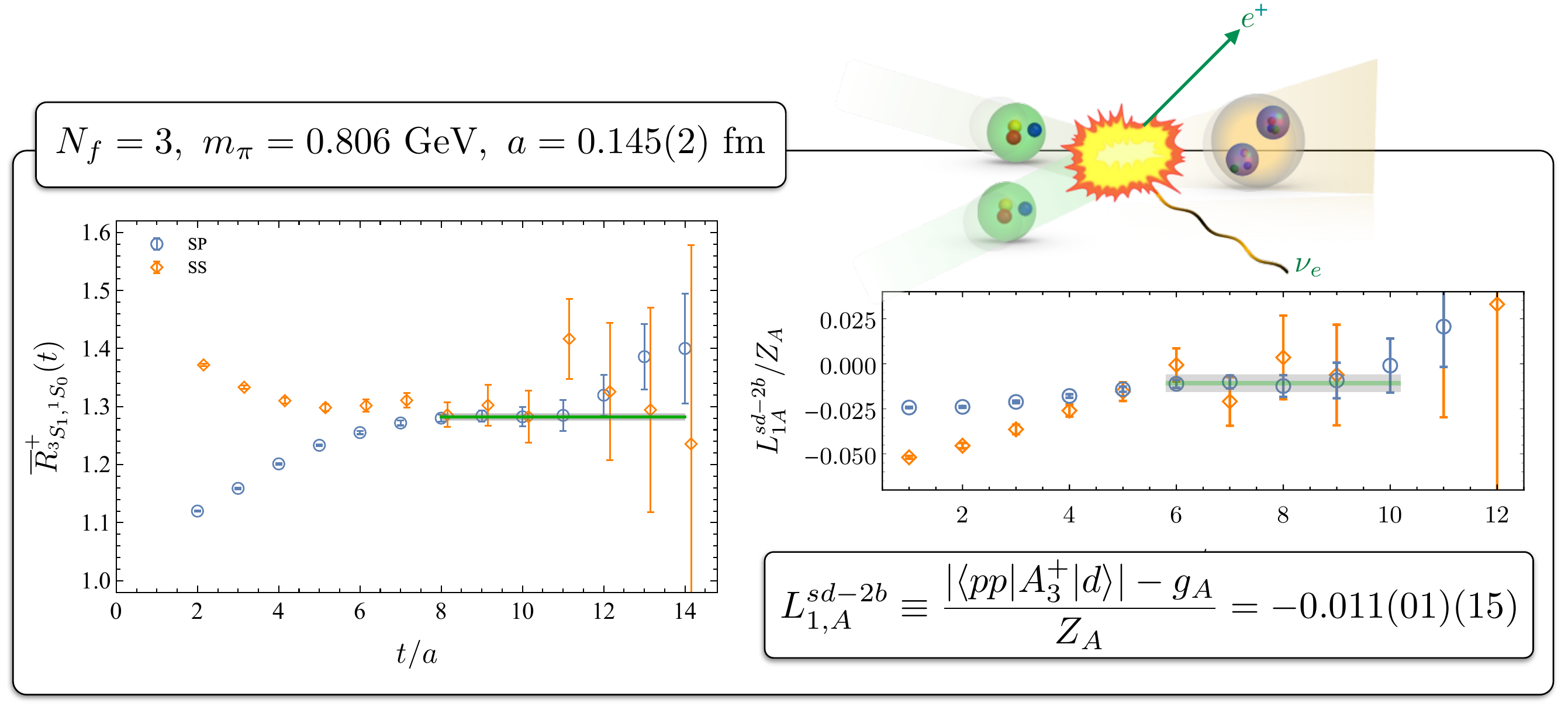}
\caption[.]{The plot in the left obtains the nuclear matrix element corresponding to the $pp$ fusion cross section obtained from LQCD with two (blue and orange) different sink operators in forming the correlation functions. The plot in the right obtains the short-distance solely two-body coupling of the pionless EFT, $L_{1,A}^{sd-2b}$. For a detailed description of quantities, see Ref.~\cite{Savage:2016kon}.}
\label{fig:pp}
\end{figure}

\noindent
While constraining nuclear reaction cross sections appeared to be a distant goal in the early stages of the field, some phenomenologically important reactions in the two-nucleon sector have now been computed from LQCD, albeit at unphysically large values of the quark masses. The first reaction studied is the M1 transition rate in the radiative capture process $np \to d \gamma$, a primary reaction in big-bang nucleosynthesis, and responsible for forming most of the light nuclei in the cosmos. An important quantity is the two-nucleon short-distance coupling of the pionless EFT, namely $L_1$, which quantifies the size of contributions to the rate beyond that induced by the magnetic moment of each of the nucleons. This coupling was constrained by applying an external constant magnetic field in a LQCD calculation to induce a transition between the two-nucleon isosinglet and isotriplet channels, at two unphysically large values of the quark masses, and was extrapolated to the physical values of the quark masses, giving rise to a cross section consistent with the experimental value~\cite{Beane:2015yha}. Perhaps a more phenomenologically interesting and less-known cross section is that of the weak counterpart of the process above. This is the $pp$ fusion process at low incident velocities, which is relevant for the energy production in sun. Here, there remains a large uncertainty on the value of the two-nucleon short-distance coupling of the pionless EFT, namely $L_{1,A}$, which is planned to be reduced to the few-percent level using the MuSun experiment. This coupling was calculated in a recent LQCD calculation from a direct evaluation of the corresponding matrix element at a large value of the quark masses, see Fig~\ref{fig:pp}, and under the assumption of a mild dependence on the quark masses, was extrapolated to the physical point~\cite{Savage:2016kon}. The obtained value was found consistent with the current phenomenological value. This study will be improved in the upcoming years towards the physical values of the quark masses.

\begin{figure}[t!]
\includegraphics[scale=0.475]{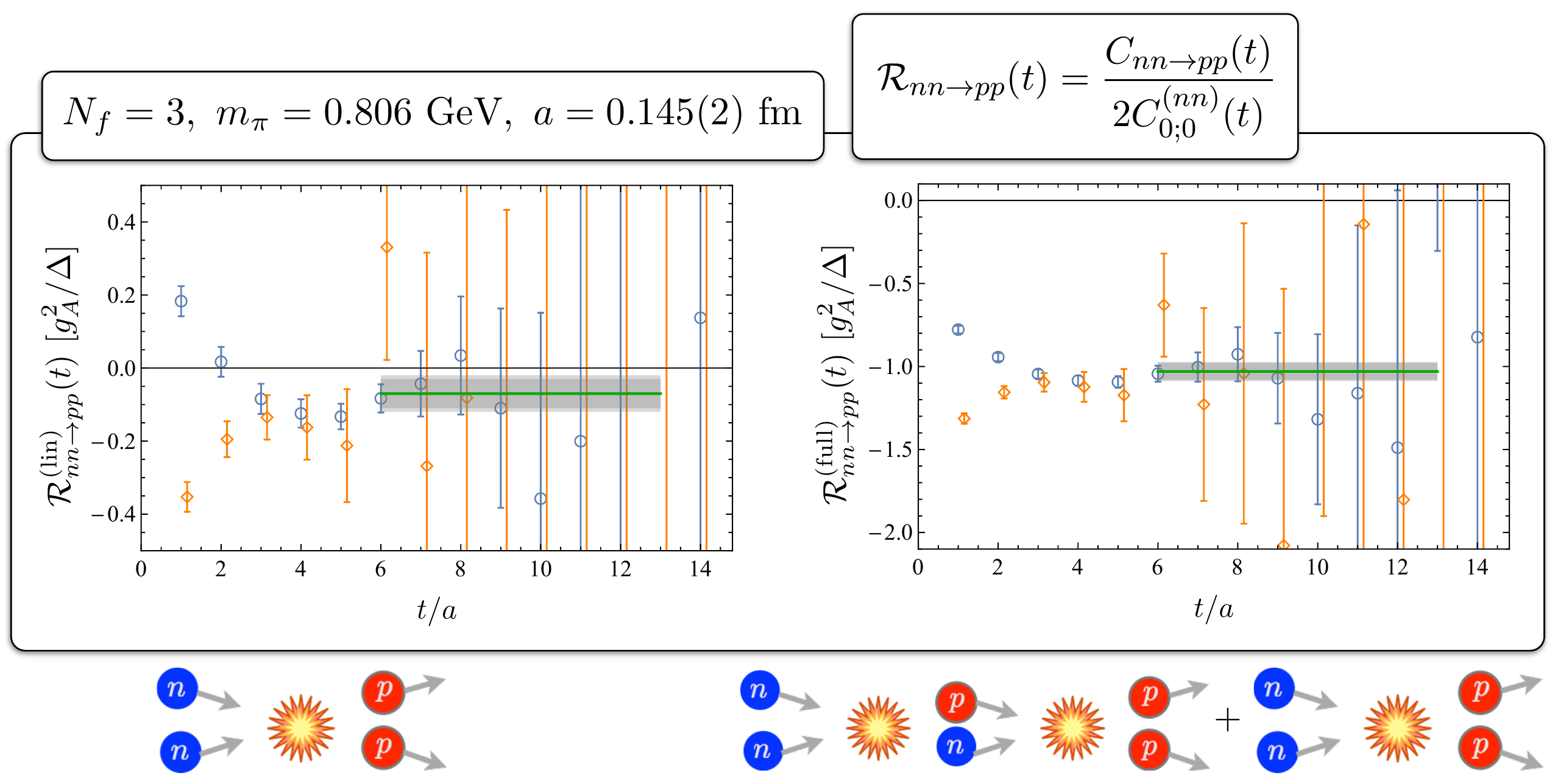}
\caption[.]{The short-distance contribution to the nuclear matrix element for $nn \to pp$ (left) and the full contribution including that from the intermediate deuteron pole (right) obtained from LQCD at larger-than-physical values of the quark masses, with two (blue and orange) different sink operators. For a detailed description of quantities, see Refs.~\cite{Shanahan:2017bgi,Tiburzi:2017iux}.}
\label{fig:nnpp}
\end{figure}

Another milestone in the area of nuclear reactions from LQCD is the study of a doubly-weak process, namely the neutrinoifull double-$\beta$ decay. In a recent study, the matrix element for the $nn \to pp$ transition was calculated, and both the long-distance contribution (arising from a deuteron pole in the intermediate state) and the short-distance contribution to the process were isolated, albeit at a large value of the quark masses, see Fig.~\ref{fig:nnpp}. A new short-distance two-nucleon doubly-weak coupling of the pionless EFT was identified for the first time and its value was constrained in this study~\cite{Shanahan:2017bgi,Tiburzi:2017iux}. It was concluded that the contribution from this coupling to the matrix element was comparable to that from the $L_{1,A}$ coupling, introducing a potential source of modification of the value of an ``effective'' axial charge in a heavy isotope. It is important to realize that such new short-distance couplings, both in the context of neutrinofull and neutrinoless double-$\beta$ decays are unknown, and in particular in the latter case, they can only be deduced from a direct LQCD calculation. This strongly motivates a continuation of this program in the upcoming years.
%

\section{Summary and Outlook
\label{sec:summary}}
\noindent
This review was aimed at showcasing a few accomplishments of the field of LQCD for nuclear physics. While most of the studies are still limited to unphysical values of the quark masses (given the significant computational cost of computations at the physical values), formal, numerical and algorithmic advancements have placed the field in a position that spectra, structure and scattering and reaction properties of few-nucleon systems, in a bound or unbound form, can be studies directly from the underlying theory. Given the emergence of exceedingly fine energy scales in the spectrum of larger nuclei and limited precision of LQCD calculations into the future, the complexity of the correlation functions when formed out of quark-level interpolating operators, and the signal-to-noise degradation in the nuclear correlation function due to the presence of lighter states in the theory, the computational complexity of LQCD-based calculations of nuclei increases dramatically with increasing the system's size. As a result, a coordinated effort by the larger community needs to be in place to systematically match experiment and LQCD results in the few-body sector to nuclear EFTs, hence enable studies of systems in the realm of nuclear many-body physics. In the meantime, with the fast and exciting progress in harnessing quantum entanglement to perform highly parallelized computations, the prospect of this alternative approach to classical computation will be investigated by nuclear physicists in the upcoming years~\cite{Carlson:2018wp}, with the ultimate goal of overcoming the impeding sign and signal-to-noise problems inherent in Monte Carlo-based studies of finite-density systems and their real-time dynamics.

\paragraph*{Acknowledgment} The USQCD collaboration acknowledges support from Nuclear Physics,  High Energy Physics and Advanced Scientific Computing Research programs at the Office of Science of the U.S. Department of Energy.

%
%

\end{document}